\documentstyle[prb,aps,twocolumn,floats,epsf]{revtex}
\def\geqap{\,\raise 2pt \hbox{$>\kern-11pt \lower 5pt \hbox{$\sim$}$}\,}
\def\leqap{\,\raise 2pt \hbox{$<\kern-10pt \lower 5pt \hbox{$\sim$}$}\,}
\begin{document}
\draft
\twocolumn[\hsize\textwidth\columnwidth\hsize\csname @twocolumnfalse\endcsname
\title{Magnetic Ordering, Orbital Ordering and Resonant X-ray Scattering in Perovskite 
Titanates}
\author{S.~Ishihara} 
\address{Department of Applied Physics, University of Tokyo, Tokyo 113-8656, Japan}
\author{T.~Hatakeyama and S.~Maekawa}
\address{Institute for Materials Research, 
Tohoku University, Sendai 980-8577, Japan}
\date{\today}
\maketitle
\begin{abstract}
The effective Hamiltonian for perovskite titanates is derived by taking into account 
the three-fold degeneracy of $t_{2g}$ orbitals and the strong electron-electron 
interactions. The magnetic and orbital ordered phases are studied in the mean-field 
approximation applied to the effective Hamiltonian. 
A large degeneracy of the orbital states in the ferromagnetic phase is found  
in contrast to the case of the doubly degenerate $e_g$ orbitals. 
Lifting of this orbital degeneracy due to lattice distortions and spin-orbit coupling is examined. 
A general form for 
the scattering cross section of the resonant x-ray scattering is derived 
and is applied to the recent experimental results in YTiO$_3$. 
The spin wave dispersion relation in the orbital ordered YTiO$_3$ is also studied. 
\end{abstract}
\pacs{PACS numbers: 75.10.-b, 71.10.-w, 78.70.-g, 71.30.+h} 
]
\narrowtext
%
%\noindent
%
\section{introdcution}
Since the discovery of layered superconducting cuprates La$_{2-x}$Sr$_x$CuO$_4$, 
studies of electronic structures of transition-metal oxides 
are revived from the modern view point of electron correlation. \cite{tokura} 
Perovskite titanates $R_{1-x}A_{x}$TiO$_3$ are one of the prototypical 
three-dimensional materials which show the Mott transition 
and anomalous metallic states at a vicinity of the transition.  
Here, $R$ and $A$ indicate the trivalent and divalent cations, respectively. 
The end compounds $R$TiO$_3$, 
where a nominal valence of all Ti ions is 3+,  
are recognized to be Mott insulators. 
A mismatch of the ionic radius in the pseudo-cubic structure 
induces the GdFeO$_3$-type lattice distortion, i.e. a tilting of a TiO$_6$ octahedron. \cite{maclean79}  
The electronic structure of $R$TiO$_3$ 
systematically changes with a kind of the rare-earth ion $R$; \cite{tokura,goral,maclean81}
a large distorted YTiO$_3$ shows a ferromagnetic ordering at 29K. 
The saturated magnetic moment 
is 0.84, \cite{garrett} which is close to the expected value from $S=1/2$, and 
a definite optical gap is observed to be about 2eV. \cite{arima,taguchi} 
On the contrary, the insulating character of 
the less distorted LaTiO$_3$ is more marginal than YTiO$_3$;   
a staggered magnetic moment for the G-type antiferromagnetic (AF) 
state is less than one half of the expected value \cite{goral} and 
an insulating gap is smaller than 0.5eV. \cite{arima,fujimori} 
By doping $R$TiO$_3$ with holes, 
the system undergoes the metal-insulator transition  
and exhibits several unconventional metallic characters  
ascribed to the electron correlation, \cite{tokura93,tokura93b,katsufuji} as well as 
La$_{2-x}$Sr$_x$CuO$_4$. 

One of the remarkable discrepancies in perovskite titanates from 
layered superconducting cuprates 
is that the orbital degree of freedom survives in titanates;    
the electron configuration of Ti$^{3+}$ 
is $d^1$ where the three $t_{2g}$ orbitals, i.e. $d_{xy}$, $d_{yz}$ and $d_{zx}$ orbitals,  
are degenerate under the cubic crystalline field. 
Thus, this ion has a degree of freedom which indicates an occupied orbital by an electron. 
The intensive and extensive studies of the orbital degree of freedom have been 
carried out recently in colossal magnetoresistive (CMR) manganites \cite{nagaosa}
where a Mn$^{3+}$ ion has doubly degenerate $e_g$ orbitals. 
Here, it is widely believed  
that observed dramatic phenomena such as CMR   
are caused by strong interplay between spin, charge and orbital as well as lattice. 
In comparison with manganites, 
there exist following characteristics in the orbital degree of freedom in titanates: 
(1) there is a three-fold degeneracy of the $t_{2g}$ orbitals, 
(2) an electron hopping between nearest-neighboring (NN)  
different orbitals is prohibited in a cubic crystal structure, 
(3) the spin-orbit (LS) coupling is possible to be relevant, 
and 
(4) the cooperative Jahn-Teller (JT) effect is weak. \cite{maclean79} 
Actually, some theoretical and experimental studies of perovskite titanates 
have been done from the view point of 
the orbital degree of freedom. \cite{mizokawa,sawada,okubo,itoh,akimitsu,keimer,khaliullin,imada,imada2,nakao} 
In addition, the resonant x-ray scattering (RXS) was applied to $R$TiO$_3$ very recently \cite{keimer,nakao}
and the orbital ordering was successfully observed in YTiO$_3$. \cite{nakao} 
The systematic studies by utilizing this experimental method are expected to clarify roles of the orbital degree of freedom 
in the Mott transition and the several unconventional phenomena observed in titanates. 

In this paper, 
the effective Hamiltonian for the electronic structures in perovskite titanates is derived  
and the spin and orbital structures in $R$TiO$_3$ are studied in the mean field approximation. 
The derived Hamiltonian corresponds to the $tJ$ model in superconducting cuprates \cite{zhang}
and the spin-orbital model for CMR manganites, \cite{ishihara}  
and is  applicable to a wide range of doped and undoped titanates. 
In particular, we focus on roles of the three-fold degeneracy in the $t_{2g}$ orbitals. 
It is shown that the orbital degeneracy of the ground state is 
more significant than that in the $e_g$ orbital case. 
A general form for 
the scattering cross section of RXS is derived 
and is applied to the recent experimental results in YTiO$_3$. \cite{nakao} 
Roles of the orbital on the spin wave dispersion relation in YTiO$_3$ are also discussed. 

\section{Effective Hamiltonian}
\label{sec:hamiltonian}
We start with the tight-binding Hamiltonian in a three dimensional lattice 
consisting of Ti ions. 
Three $t_{2g}$ orbitals $d_{xy}$, $d_{yz}$ and $d_{zx}$ 
and the intra-atomic Coulomb interactions are considered in each Ti ion. 
The Hamiltonian is 
\begin{eqnarray}
{\cal H}
&=&\sum_{\langle i j \rangle \gamma \gamma' \sigma} \Bigl (t_{ij}^{\gamma \gamma'}
 d_{i \gamma \sigma}^\dagger d_{j \gamma' \sigma} +H.c. \Bigr )\nonumber \\
&+&U\sum_{i \gamma} n_{i \gamma \uparrow} n_{i \gamma \downarrow}
+U'{1 \over 2}\sum_{i \gamma \ne \gamma'} n_{i \gamma} n_{i \gamma'}
\nonumber \\
&+& I\sum_{i \gamma > \gamma' \sigma \sigma'} 
d_{i \gamma \sigma}^\dagger d_{i \gamma' \sigma'}^\dagger
d_{i \gamma \sigma'}        d_{i \gamma' \sigma} 
\nonumber \\
&+& I \sum_{i \gamma \ne \gamma'} 
d_{i \gamma  \uparrow}^\dagger d_{i \gamma  \downarrow}^\dagger
d_{i \gamma' \downarrow}  d_{i \gamma' \uparrow}         , 
\label{eq:humorig}
\end{eqnarray}
where $d_{i \gamma \sigma}^\dagger$ creates a $t_{2g}$ electron at site $i$ 
with spin $\sigma(=\uparrow, \downarrow)$ 
and orbital $\gamma(=xy,yz,zx)$.
$U$ and $U'$ are the intra-orbital and inter-orbital Coulomb interactions, respectively, 
and $I$ is the exchange interaction. 
In an isolated ion, 
these interactions are represented by the Racah parameters as 
$U=A+4B+3C$, $U'=A-2B+C$ and $I=3B+C$, 
and a relation $U=U'+2I$ is satisfied. 
$t_{ij}^{\gamma \gamma'}$ is the hopping integral between site $i$ and its 
NN site $j$ with orbitals $\gamma$ and $\gamma'$, respectively. 
In a simple cubic lattice, 
the hopping integral is diagonal  
and one of the diagonal components is zero. 
For a Ti-Ti bond in a direction $l(=x,y,z)$, 
two orbitals, which have a finite hopping integral $t_{ij}^{\gamma \gamma}$,  
are termed active orbitals denoted by $a_l$ and $b_l$, 
and one with no hopping integral is termed an inactive orbital denoted by $c_l$. 
For example, $(a_x, b_x, c_x)=(zx,xy,yz)$. 
$t_{ij}^{\gamma \gamma'}$ is simply expressed in this case as 
\begin{equation}
t_{ij}^{\gamma \gamma'}=t \delta_{\gamma \gamma'} (\delta_{\gamma a_l}+\delta_{\gamma b_l}) . 
\label{eq:ttt}
\end{equation}
The GeFeO$_3$-type lattice distortion breaks this relation  
as discussed in Sec.~\ref{sec:so}. 

Since the Coulomb interactions $U$ and $U'$ are larger than the hopping integral in titanates, \cite{bocquet} 
the effective Hamiltonian is derived by perturbational calculation with respect to the hopping integral. 
The Hamiltonian is  
\begin{equation}
{\cal H}={\cal H}_{t}+{\cal H}_{J},  
\label{eq:effective}
\end{equation}
where the first and second terms correspond to the so-called $t$ and $J$ terms in the $tJ$ model, respectively. 
The $t$ term is given by 
\begin{equation}
{\cal H}_t=\sum_{\langle i j \rangle \gamma \gamma' \sigma} t_{ij}^{\gamma \gamma'}
{\widetilde d}_{i \gamma \sigma}^\dagger {\widetilde d}_{j \gamma' \sigma} +H.c. , 
\label{eq:ht}
\end{equation}
where ${\widetilde d}_{i \gamma \sigma} (=d_{i \gamma \sigma}
\Pi_{(\gamma' \sigma') \ne (\gamma \sigma)}(1-d_{i \gamma' \sigma'}^\dagger d_{i \gamma' \sigma'}))$ 
excludes multi occupied states of electrons at site $i$. 
The $J$ term is classified by the point symmetry of the intermediate electronic states, i.e. $d^2$ states. 
In the case where Eq.~(\ref{eq:ttt}) is satisfied, 
the Hamiltonian is 
\begin{eqnarray}
{\cal H}_J={\cal H}_{T_1}+{\cal H}_{T_2}+{\cal H}_{E}+{\cal H}_{A_1}, 
\label{eq:eff}
\end{eqnarray}
with 
\begin{equation}
{\cal H}_{T_1}=-J_{T_1} \sum_{\langle ij \rangle}
\biggl ( {3 \over 4}n_i n_j+\vec S_i \cdot \vec S_j \biggr) 
\Big (  B^l-C^l+D^l    \Bigr ) , 
\label{eq:ht1}
\end{equation}
\begin{equation}
{\cal H}_{T_2}=-J_{T_2} \sum_{\langle ij \rangle}
\biggl ( {1 \over 4}n_i n_j-\vec S_i \cdot \vec S_j \biggr) 
\Big ( B^l+C^l+D^l    \Bigr ) , 
\label{eq:ht2}
\end{equation} 
\begin{equation}
{\cal H}_{E}=-J_{E} \sum_{\langle ij \rangle}
\biggl ( {1 \over 4}n_i n_j-\vec S_i \cdot \vec S_j \biggr) 
\biggl (  {2 \over 3} A^l-{2 \over 3} C'^l \biggr)    , 
\label{eq:he}
\end{equation}
\begin{equation}
{\cal H}_{A_1}=-J_{A_1} \sum_{\langle ij \rangle}
\biggl ( {1 \over 4}n_i n_j-\vec S_i \cdot \vec S_j \biggr) 
\biggl ( {1 \over 3} A^l +{2 \over 3} C'^l \biggr)    .  
\label{eq:ha1}
\end{equation}
Prefactors are given by 
$J_{T_1}=t^2/(U'-I)$, $J_{T_2}=t^2/(U'+I)$, $J_{E}=t^2/(U-I)$ and $J_{A_1}=t^2/(U+2I)$. 
By using the relation $U=U'+2I$, 
we obtain $J_{T_2}=J_{E}$. 
$n_i(=\sum_{\sigma \gamma}d_{i \gamma \sigma}^\dagger d_{i \gamma \sigma})$ 
is the number operator 
and 
$\vec S_i$ is the spin operator given by  
\begin{equation}
\vec S_i={1 \over 2} \sum_{\gamma \sigma \sigma'}  
d^\dagger_{i \gamma \sigma} \vec \sigma_{\sigma \sigma'} d_{i \gamma \sigma'} . 
\label{eq:spin}
\end{equation}
$A^l$, $B^l$, $C^l$, $C'^l$ and $D^l$ are the orbital parts of the Hamiltonian 
represented by the eight orbital operators $O_{\Gamma \gamma}$ 
where 
$\Gamma$ denotes an irreducible representation in the $O_h$ group 
and $\gamma$ classifies the bases of the irreducible representation.
To represent $O_{\Gamma \gamma}$, let us introduce 
the Gell-Mann matrices which are generators of the SU(3) algebra: \cite{gellmann}  
\begin{equation}
\lambda_{1}=\left ( 
   \begin{array}{@{\,} ccc @{\,}}
   0 & 1 & 0 \\ 
   1 & 0 & 0 \\
   0 & 0 & 0
   \end{array}
   \right ) , 
\qquad
\lambda_{2}=\left ( 
   \begin{array}{@{\,} ccc @{\,}}
   0 & -i & 0 \\ 
   i & 0 & 0 \\
   0 & 0 & 0
   \end{array}
   \right )  ,
\end{equation}
\begin{equation}
\lambda_{3}=\left ( 
   \begin{array}{@{\,} ccc @{\,}}
   -1 & 0 & 0 \\ 
   0 & 1 & 0 \\
   0 & 0 & 0
   \end{array}
   \right )  ,
\qquad   
   \lambda_{4}=\left ( 
   \begin{array}{@{\,} ccc @{\,}}
   0 & 0 & 1 \\ 
   0 & 0 & 0 \\
   1 & 0 & 0
   \end{array}
   \right ) , 
\label{eq:oe}
\end{equation}
\begin{equation}
\lambda_{5}=\left ( 
   \begin{array}{@{\,} ccc @{\,}}
   0 & 0 & i \\ 
   0 & 0 & 0 \\
   -i & 0 & 0
   \end{array}
   \right ) , 
\qquad
\lambda_{6}=\left ( 
   \begin{array}{@{\,} ccc @{\,}}
   0 & 0 & 0 \\ 
   0 & 0 & 1 \\
   0 & 1 & 0
   \end{array}
   \right ) ,
\end{equation}
\begin{equation}
\lambda_{7}=\left ( 
   \begin{array}{@{\,} ccc @{\,}}
   0 & 0 & 0 \\ 
   0 & 0 & -i \\
   0 & i & 0
   \end{array}
   \right ) ,
\qquad 
\lambda_{8}={1 \over \sqrt{3}}\left ( 
   \begin{array}{@{\,} ccc @{\,}}
   1 & 0 & 0 \\ 
   0  & 1 & 0 \\
   0 & 0 & -2
   \end{array}
   \right ) . 
\label{eq:ot1} 
\end{equation}
We define the orbital operators as 
\begin{equation}
O_{i \Gamma \gamma}={-1 \over \sqrt{2}} \sum_{\sigma \alpha \beta} 
d_{i \alpha \sigma}^\dagger (\lambda_l)_{\alpha \beta} d_{i \beta \sigma} , 
\label{eq:orbop}
\end{equation}
where 
$(\Gamma \gamma; l)=$$(Eu; 8)$, $(Ev, 3)$, 
$(T_2 x; 6)$, $(T_2 y; 4)$, $(T_2 z; 1)$, 
$(T_1 x; 7)$, $(T_1 y; 5)$ and $(T_1 z; 2)$.  
The operators 
$O_{i E \gamma}$ and $O_{i T_2 \gamma}$ describe the electric quadrupole moments and 
$O_{i T_1 \gamma}$ describes the magnetic dipole moment. 
By utilizing the above operators, 
we obtain 
\begin{eqnarray}
A^l&=&2 \biggl ({2 \over 3}-\sqrt{2 \over 3}O^l_{i Eu}   \biggr ) 
                   \biggl ({2 \over 3}-\sqrt{2 \over 3}O^l_{j Eu}   \biggr )
                 \nonumber \\
   &+&4 O^l_{i Ev} O^l_{j Ev} , 
\label{eq:aaa}
\end{eqnarray}
\begin{eqnarray}
B^l&=& \biggl ({2 \over 3}-\sqrt{2 \over 3}O^l_{i Eu}   \biggr ) 
       \biggl ({2 \over 3}-\sqrt{2 \over 3}O^l_{j Eu}   \biggr )
\nonumber \\
   &-&2 O^l_{i Ev} O^l_{j Ev} , 
\label{eq:bbb}
\end{eqnarray}
\begin{equation}
C^l=2 \bigl (O_{i T_2 l} O_{j T_2 l}+O_{i T_1 l}O_{j T_1 l} \bigr), 
\label{eq:ccc}
\end{equation}
\begin{equation}
C^{'l}=2(O_{i T_2 l} O_{j T_2 l}-O_{i T_1 l}O_{j T_1 l} ), 
\label{eq:ccc2}
\end{equation}
\begin{eqnarray}
D^l&=&\biggl( {1 \over 3} +\sqrt{2 \over 3} O^l_{i Eu} \biggr )
    \biggl( {2 \over 3} -\sqrt{2 \over 3} O^l_{j Eu} \biggr )
    \nonumber \\
   &+&\biggl( {2 \over 3} -\sqrt{2 \over 3} O^l_{i Eu} \biggr )
    \biggl( {1 \over 3} +\sqrt{2 \over 3} O^l_{j Eu} \biggr ) ,   
\label{eq:ddd}
\end{eqnarray}
where $O^l_{i E \gamma}$'s are given by 
\begin{equation}
\left ( 
   \begin{array}{@{\,} c @{\,}}
   O^l_{i Eu} \\ 
   O^l_{i Ev}  
\end{array}
\right ) 
 =
\left ( 
   \begin{array}{@{\,} cc @{\,}}
    \cos {2 \pi \over 3}m_l & \sin{2 \pi \over 3}m_l  \\
   -\sin {2 \pi \over 3}m_l & \cos{2 \pi \over 3}m_l 
   \end{array}
   \right ) 
\left ( 
   \begin{array}{@{\,} c @{\,}}
   O_{i Eu} \\ 
   O_{i Ev}  
\end{array}
\right )   ,
\end{equation}
%
%fig1
\begin{figure}
\epsfxsize=0.7\columnwidth
\centerline{\epsffile{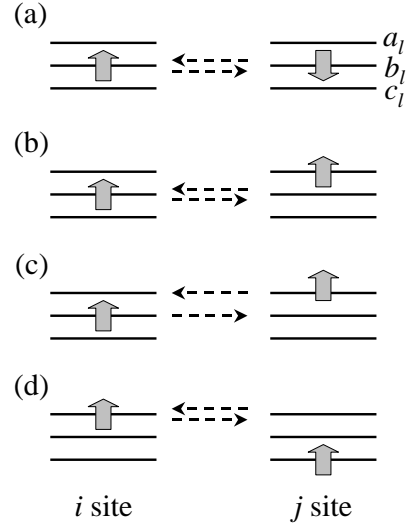}}
\caption{
Schematic pictures of representative virtual exchange processes 
for the terms 
(a) $ A^l$, (b) $B^l$, (c) $C^l$ and (d) $D^l$ defined in Eqs.~(\ref{eq:aaa}), 
(\ref{eq:bbb}), (\ref{eq:ccc}) and (\ref{eq:ddd}), respectively. 
$a_l$ and $b_l$ are the active orbitals and $c_l$ is the inactive orbital 
(see text).  
}
\label{fig:process}
\end{figure}
with $m_l=(1,2,3)$ for $l=(x,y,z)$. 
It is worth rewriting the orbital parts of the Hamiltonian from the view point of 
the active and inactive orbitals; we introduce the pseudo-spin operator with 
the quantum number $1/2$ for a bond along a direction $l$ 
\begin{equation}
\vec T_i^l={1 \over 2} \sum_{\gamma, \gamma'=(a_l, b_l), \sigma}
d_{i \gamma \sigma}^\dagger \vec \sigma_{\gamma \gamma'} d_{i \gamma' \sigma} , 
\label{eq:pspsps}
\end{equation}
with $\vec \sigma$ being the Pauli matrices, 
and the number operator 
$n_{i \gamma_l}=\sum_{\sigma} d_{i \gamma_l \sigma}^\dagger   d_{i \gamma_l \sigma}$ 
for $\gamma_l=(a_l, b_l, c_l)$. 
The orbital parts of the Hamiltonian 
are rewritten as 
\begin{eqnarray}
A^l&=&4\bigl (n_{i a_l}n_{j a_l}+n_{i b_l}n_{j b_l} \bigr) , \\
B^l&=&2\bigl (n_{i a_l}n_{j b_l}+n_{i b_l}n_{j a_l} \bigr) , \\
C^l&=&4\bigl (T_{i x}^l T_{j x}^l+T_{i y}^lT_{j y}^l  \bigr) , \\
C^{'l}&=&4 \bigl (T_{ix}^l T_{jx}^l-T_{iy}^l T_{jy}^l \bigr), \\ 
D^l&=&n_{i c_l} \bigl (n_{j a_l}+n_{j b_l} \bigr )+\bigl (n_{i a_l}+n_{i b_l} \bigr)n_{j c_l} . 
\end{eqnarray} 
We note that $\vec T_{i}^l$'s  with different $l$ are not independent with each other. 
These simple expressions result from the diagonal form in the hopping integral under which 
an electron number of each orbital is conserved. 
Schematic pictures of representative exchange processes for $A^l$, $B^l$, $C^l$ and $D^l$ 
are shown in Fig.~\ref{fig:process}. 

A similar model Hamiltonian with Eq.~(\ref{eq:eff}) was derived in 
Ref.~\onlinecite{kugel}, although the intra-atomic exchange interaction $I$ is assumed to be zero. 
As shown in the next section, 
this condition corresponds to a critical point in the phase diagram. 
Also, a similar model with $I=0$ is represented by 
the pseudo-spin operators $T_{i}^l$ (Eq.~(\ref{eq:pspsps})) 
in Ref.~\onlinecite{khaliullin}. In a model Hamiltonian derived in Ref.~\onlinecite{imada}, 
two of the three $t_{2g}$ orbitals are taken into account.  

\section{Spin and Orbital States}
\label{sec:so}
The effective Hamiltonian is applied to $R$TiO$_3$  
where a valence of all Ti ion is $3+$. 
Consider a hypothetical-cubic lattice consisting of Ti ions, 
instead of the actual crystal lattice of $R$TiO$_3$.  
The mean-field approximation at zero temperature is applied to the Hamiltonian and 
the four kinds of spin and orbital ordered states are considered; 
a uniform spin (orbital) state termed F  
and three staggered spin (orbital) states termed A-AF, C-AF and G-AF. 
The periodicities of these orderings are characterized by the momentum 
$(0 0 0)$, $(0 0 \pi)$, $(\pi \pi 0)$, and $(\pi \pi \pi)$, respectively. 
$\langle S_{i z} \rangle(=\pm 1/2)$ is adopted to be a spin order parameter 
and, the orbital order parameters $\langle O_{i \Gamma \gamma} \rangle$ are calculated 
by the orbital wave function at site $i$  
\begin{equation}
| \psi_i \rangle=C_{ixy}| d_{ixy}\rangle+C_{iyz}| d_{iyz}\rangle+C_{izx}| d_{izx}\rangle . 
\label{eq:ws}
\end{equation}
The order parameters are optimized numerically 
to obtain the lowest energy. 
%
%figure2
\begin{figure}
\epsfxsize=0.9\columnwidth
\centerline{\epsffile{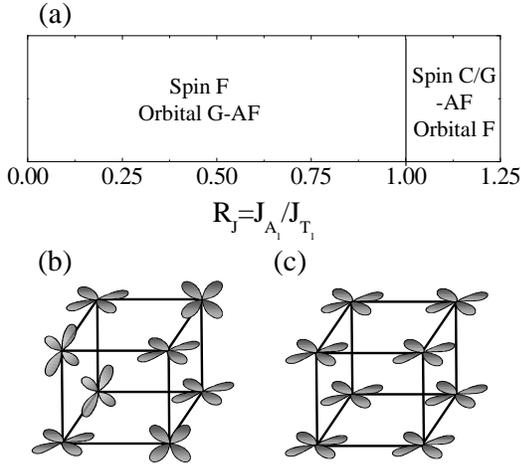}}
\caption{
(a): The spin and orbital phase diagram as a function of $R_J(\equiv J_{A_1}/J_{T_1})$. 
Schematic pictures of the representative orbital ordered states for $R_J<1$ and 
$R_J>1$ are shown in (b) and (c), respectively. 
}
\label{fig:cubicps}
\end{figure}
In Fig.~\ref{fig:cubicps}, 
the magnetic and orbital phase diagram is presented 
as a function of $R_J\equiv J_{A_1}/J_{T_1}$. 
The relation $J_{T_2}/J_{T_1}=5R_J/(2+3R_J)$ derived from  
the condition $U=U'+2I$ is used. 
Although a value of $R_J$ is smaller than one in actual compounds, 
the calculated results in the region of $R_J>1$ are also shown for comparison. 
In the region of $R_J<1$, 
the F spin state 
and the G-AF orbital ordered state are realized.
The wave functions in the $A$ and $B$ orbital sublattices are
given by 
\begin{eqnarray}
|\psi_A \rangle&=&|d_{\alpha} \rangle  ,  \nonumber \\
|\psi_B \rangle&=& \cos\theta | d_{\beta} \rangle +\sin \theta e^{i \phi} | d_{\gamma} \rangle ,  
\label{eq:waab}
\end{eqnarray}
respectively, 
with any values of $\theta \in[0 , \pi]$ and $\phi \in [0 , 2\pi]$, and 
$(\alpha, \beta, \gamma)=(xy,yz,zx)$, $(yz,zx,xy)$, $(zx,xy,yz)$. 
There is a large continuous degeneracy in the orbital state; 
any linear combinations of $|d_{\beta} \rangle$ and $|d_{\gamma} \rangle$ are degenerate in the sublattice $B$. 
This is generalized to the states 
\begin{eqnarray}
|\psi_{i_A} \rangle&=&|d_{\alpha} \rangle  ,  \nonumber \\
|\psi_{j_B} \rangle&=& \cos\theta_{j_B} | d_{\beta} \rangle +\sin \theta_{j_B} e^{i \phi_{j_B}} | d_{\gamma} \rangle ,  
\label{eq:waabg}
\end{eqnarray}
where $i_A$ ($j_B$) indicates the $i(j)$-th site in the $A(B)$ orbital sublattice. 
$\theta_{j_B}$ and $\phi_{j_B}$ at each site are taken independently, because the Hamiltonian 
includes the interactions between the NN sites.
In order to understand this state in more detail, 
let us consider the hopping integral 
between the NN occupied orbitals defined by 
\begin{equation}
\tau_{ij}=\Bigl \langle \psi_i \Bigl | 
\sum_{<ij> \gamma \gamma' \sigma}
t_{ij}^{\gamma \gamma'} {\widetilde d}_{i \gamma \sigma}^\dagger {\widetilde d}_{j \gamma' \sigma} +H.c. 
\Bigr |\psi_j \Bigr \rangle . 
\label{eq:tau}
\end{equation}
$\tau_{ij}$'s are zero for all bonds in this orbital ordered state. 
Therefore, the exchange processes denoted by 
$A^l$ (Eq.~(\ref{eq:aaa})) do not occur (see Fig.~\ref{fig:process}), 
and the ferromagnetic interaction is dominant. 
It is worth comparing the present results with 
those in the system where the doubly degenerate $e_g$ orbitals, 
i.e. the $d_{3z^2-r^2}$ and $d_{x^2-y^2}$ orbitals, exist. 
Here, the effective Hamiltonian corresponding to Eq.~(\ref{eq:eff}) 
is expressed by the spin operator $\vec S_i $ and the pseudo-spin operator 
for the orbital degree of freedom $\vec T_{i}$ with the quantum number $1/2$. \cite{ishihara,kugel2} 
The orbital state in the ferromagnetic phase obtained by the mean-field approximation 
is the G-AF orbital state where  
the wave functions are given by 
\begin{eqnarray}
|\psi_A \rangle&=& \cos \Bigl ( {\theta \over 2} \Bigr ) |d_{3z^2-r^2} \rangle 
                 + \sin \Bigl ( {\theta \over 2} \Bigr ) |d_{x^2-y^2} \rangle ,  \nonumber \\
|\psi_B \rangle&=&-\sin \Bigl ( {\theta \over 2} \Bigr ) |d_{3z^2-r^2} \rangle 
                 + \cos \Bigl ( {\theta \over 2} \Bigr ) |d_{x^2-y^2} \rangle , 
\label{eq:weg}
\end{eqnarray}
for any value of $\theta \in [0 , 2\pi]$.\cite{ishihara4} 
This is also a staggered orbital ordered state with a continuous degeneracy. 
However, there exist the remarkable differences 
between the $t_{2g}$ and $e_g$ cases: 
(1) The hopping integrals between the occupied orbitals 
$\tau_{ij}$ are finite in the $e_g$ case 
and depend on $\theta$. 
Thus, the generalization of the orbital ordered state, 
as seen in the $t_{2g}$ case (from Eq.~(\ref{eq:waab}) to Eq.~(\ref{eq:waabg})), is impossible.
That is, the orbital degeneracy in the ground state is more remarkable in the $t_{2g}$ case. 
This is because a number of the orbital degree of freedom is larger in this case.  
(2) The orbital wave functions with complex coefficients 
are not included in the $e_{g}$ case.  
This is because the effective Hamiltonian for the $e_g$ electron 
is represented by the operators $T_x$ and $T_z$, 
unlike $T_y$ which breaks the time reversal symmetry. 
In the region of $R_J>1$ in Fig.~\ref{fig:cubicps},  
there is no continuous orbital degeneracy.  
The C-AF and G-AF spin states are realized associated with    
the F orbital state with $|\psi \rangle=|d_{\alpha}\rangle$ for $\alpha=(xy, yz, zx)$. 
The point $R_J=1$, i.e. $I=0$,  
is a critical point in the phase diagram where 
degeneracy of the spin and orbital states are significant. 
It is supposed that 
$R_J$ for $R$TiO$_3$ is about $0.2 \sim 0.4$,\cite{mizokawa} although 
there is a large ambiguity in an estimation of $I$.

%
%figure3
\begin{figure}
\epsfxsize=0.6\columnwidth
\centerline{\epsffile{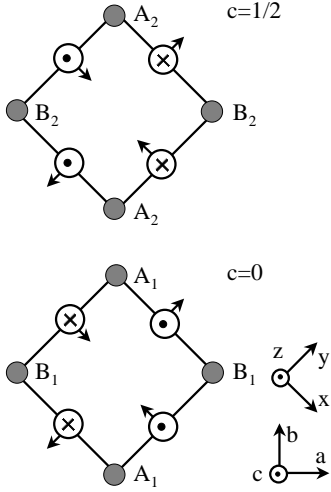}}
\caption{
A schematic picture of the crystal structure of $R$TiO$_3$. 
Filled and open circles indicate Ti and O ions, respectively, 
and $A_1$, $A_2$, $B_1$ and $B_2$ denote Ti sites in a unit cell. 
Arrows indicate displacements of O ions in the $ab$ plane, 
and crosses and dots in the O ions indicate the displacement along the $+c$ and $-c$ directions, 
respectively. 
}
\label{fig:crystal}
\end{figure}
We next examine how this large continuous orbital degeneracy is lifted  
by the following three effects observed in $R$TiO$_3$: 
the GdFeO$_3$-type lattice distortion, 
the JT-type distortion in a TiO$_6$ octahedron and the LS coupling.  
The GdFeO$_3$-type lattice distortion bends a Ti-O-Ti bond. 
The simple form of the hopping integral in Eq.~(\ref{eq:ttt}) is not valid 
and $t_{ij}^{\gamma \gamma'}$ for any pairs of $\gamma$ and $\gamma'$ are finite. 
We calculate all components of the hopping integral $t_{ij}^{\gamma \gamma'}$ 
for the crystal structures in  YTiO$_3$ and LaTiO$_3$ 
by the Slater-Koster formula. \cite{slater} 
The most remarkable changes are found in 
the hopping integrals between the different active orbitals $t_{ij}^{a_l b_l}$.
This is because 
the GdFeO$_3$-type distortion induces a $\sigma$ bond between 
the $d_{a_l}$ ($d_{b_l}$) orbital and the $2p_{l}$ orbital 
at the NN O site. 
We simulate this distortion by introducing a new term in the hopping integral as 
\begin{equation}
t_{ij}^{\gamma \gamma'}=(t_\pi     \delta_{\gamma \gamma'}+s_l t_\sigma  \delta_{\gamma \ne \gamma'} )  
(\delta_{\gamma a_l}+\delta_{\gamma b_l}) ,  
\label{eq:pisigma}
\end{equation}
where the sign of the tranfer integral $s_l=\pm1$ depends on the direction $l$ 
and a parameter $R_t=t_\sigma/t_\pi$ is interpreted to be an increasing function of 
this distortion. 
A value of $R_t$ for YTiO$_3$ is estimated to be about $0.5 \sim 1$. 
The effective Hamiltonian  
including the GdFeO$_3$-type distortion, the JT-type distortion and the LS coupling is 
given by 
\begin{equation}
{\cal H}={\widetilde {\cal H}}_J+{\cal H}_{JT}+{\cal H}_{LS}, 
\label{eq:effective2}
\end{equation}
where $t$ term is neglected. 
The first term is the modified $J$ term including 
the GdFeO$_3$-type distortion and its explicit form is presented in the Appendix \ref{sec:ap1}. 
The second term is for the JT coupling 
\begin{eqnarray}
{\cal H}_{JT}&=&
 g_{E}   \sum_{i \gamma=u,v} O_{i E \gamma} Q_{i E \gamma}
 \nonumber \\
&+&g_{T_2} \sum_{i \gamma=x,y,z} O_{i T_2 \gamma} Q_{i T_2 \gamma} ,  
\label{eq:jt}
\end{eqnarray}
where $Q_{\Gamma \gamma}$'s 
are the normal coordinates of the oxygen displacements in a TiO$_6$ octahedron 
with the symmetry $\Gamma \gamma$, and $g_{\Gamma}$'s are the coupling constants. 
The last term is for the LS coupling denoted by 
\begin{equation}
{\cal H}_{LS}=\lambda_{LS} \sum_{i \gamma=x, y, z} S_{i \gamma} O_{i T_1 \gamma}. 
\label{eq:ls}
\end{equation}
The corrections for the effective Hamiltonian in Eq.~(\ref{eq:effective2}) 
are of the order of 
$O(g_{\Gamma}t^2/U^2)$ and $O(\lambda_{LS} t^2/U^2)$. 
In the mean field calculation, 
an orthorhombic unit cell with the $Pbnm$ space group is adopted. 
Four Ti ions in a unit cell are termed 
$A_1$, $A_2$, $B_1$ and $B_2$ (see Fig.~\ref{fig:crystal})  
and the orbital states at these sites are considered independently.
As for the JT-type distortion, 
$Q_{T_2 y}$ and ${- 1 \over 2}Q_{E u}+{\sqrt{3} \over 2}Q_{E v}(\equiv Q_{E 3x^2-r^2}) $ 
are dominant at site $A_1$ in YTiO$_3$. 
Thus, we assume in this calculation that 
$|g_{T_2}| Q_{A_1 T_2 y}=|g_{E }|Q_{A_1 E 3x^2-r^2} (\equiv gQ)$ 
and other components are zero. 
The JT-type distortions at other sites are introduced by considering the $Pbnm$ space group. 

%
%figure4
\begin{figure}
\epsfxsize=0.7\columnwidth
\centerline{\epsffile{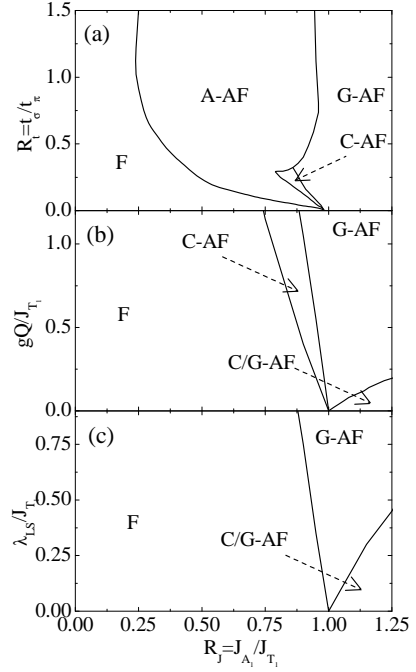}}
\caption{
The magnetic phase diagrams as functions of $R_J$ and 
(a) a ratio between the hopping integrals $R_t=t_{\sigma}/t_{\pi}$ caused by the 
GdFeO$_3$-type lattice distortion, 
(b) the JT-type lattice distortion $gQ$, 
and 
(c) LS coupling $\lambda_{LS}$. 
Other parameter values are chosen to be 
$\lambda_{LS}=R_t=0$ in (a), $gQ=\lambda_{LS}=0$ in (b), and $gQ=R_t=0$ in (c). 
}
\label{fig:psadd}
\end{figure}
The phase diagrams 
with including the GdFeO$_3$-type lattice distortion, 
the JT-type distortion and the LS coupling 
are presented in Figs.~\ref{fig:psadd} (a), (b) and (c), respectively. 
In all phase diagrams, 
the large orbital degeneracy is lifted and 
a region of the ferromagnetic phase shrinks. 
This result implies that 
the high symmetry in the orbital space  
is favorable for the ferromagnetic spin ordering. 
The characteristics of the each phase diagram are summarized as follows; 
(1) By introducing the GdFeO$_3$-type lattice distortion, 
the orbital degeneracy is partially lifted 
and one of orbital ordered states in the ferromagnetic phase is given by 
$|\psi_{A_1} \rangle= 0.71|d_{xy} \rangle+0.71|d_{yz} \rangle$, 
$|\psi_{B_1} \rangle= 0.38|d_{xy} \rangle-0.35|d_{yz} \rangle+0.85|d_{zx} \rangle$,
$|\psi_{A_2} \rangle= 0.38|d_{xy} \rangle-0.85|d_{yz} \rangle+0.35|d_{zx} \rangle$ and 
$|\psi_{B_2} \rangle=-0.71|d_{xy} \rangle+0.71|d_{zx} \rangle$. 
This lattice distortion 
brings about new terms $C^{'l}$, $D^{'l}$ and $E^l$ 
in the Hamiltonian (see Appendix \ref{sec:ap1}) 
which promote a mixing of the different kinds of orbitals.   
Thus, the uniform components of the orbital increase  
and the hopping integral between the occupied orbitals
$ \tau_{ij}$ becomes finite. 
As a result, 
the ferromagnetic ordering is relatively unstable in comparison with the AF spin ordering  
due to the term $A^l$.  
(2) 
The large JT-distortion with the $E_g$ symmetry $g_E Q_{E}$ favors the orbital ordered state of  
any linear combinations of the $d_{xy} $ and $d_{zx} $ orbitals for 
$A_1$ and $A_2$ sites, 
and those of the $d_{xy} $ and $d_{yz} $ orbitals for 
$B_1$ and $B_2$ sites. 
On the contrary, 
$g_{T_2} Q_{T_2}$ lifts the degeneracy uniquely as 
$|\psi_{A_1} \rangle={1 \over \sqrt{2}}(|d_{xy} \rangle-|d_{zx} \rangle)$, 
$|\psi_{A_2} \rangle={1 \over \sqrt{2}}(|d_{xy} \rangle+|d_{zx} \rangle)$, 
$|\psi_{B_1} \rangle={1 \over \sqrt{2}}(|d_{xy} \rangle+|d_{yz} \rangle)$ and 
$|\psi_{B_2} \rangle={1 \over \sqrt{2}}(|d_{xy} \rangle-|d_{yz} \rangle)$. 
The obtained orbital state in the F spin phase in 
Fig.~\ref{fig:psadd}(b) is the mixed state of the two. 
Here, the hopping integrals $\tau_{ij}$'s 
become finite for all NN bonds. 
Thus, a region of the ferromagnetic phase shrinks by introducing $gQ$. 
(3) The LS coupling fixes 
the direction of spins in the ferromagnetic phase along [100].  
The large orbital degeneracy at $\lambda_{LS}=0$ is lifted 
and one of the obtained orbital state is given by 
$|\psi_{A_1} \rangle=|\psi_{B_2} \rangle={1 \over \sqrt{2}}(i|d_{yz} \rangle+|d_{zx} \rangle)$ and 
$|\psi_{A_2} \rangle=|\psi_{B_1} \rangle=|d_{xy} \rangle$. 
Although this orbital ordered state is included in the solutions 
at $\lambda_{LS}=0$ shown in Eq,~(\ref{eq:waabg}),  
there is no energy gain for the LS coupling at sites $A_2$ and $B_1$. 
This is because the F spin state is not compatible with the staggered 
orbital ordered state in the large limit of $\lambda_{LS}$. 

\section{Resonant X-ray Scattering}
\label{sec:rxs}
RXS was first applied to observation of the orbital ordering in 
perovskite manganites. \cite{murakami} 
By tuning the incident x-ray energy to the $K$-edge of the transition-metal ion, 
the atomic scattering factor becomes a tensor with respect to the polarization of x ray 
and is sensitive dramatically to the local electronic structure. \cite{ishihara0}
In this section, the scattering cross section of RXS
in titanates is formulated.  
This is applied to the recent experimental results in YTiO$_3$.
Let us consider the scattering of x ray with momentum $\vec k_i$, energy $\omega_i$
and polarization $\lambda_i$ to $\vec k_f$, $\omega_f$ and $\lambda_f$. 
The electronic states at the initial, final and intermediate states of the scattering are denoted  
by $| i\rangle$, $| f \rangle$ and $|m \rangle$  
with energies $\varepsilon_{i}$, $\varepsilon_f$ and $\varepsilon_m$, respectively.  
We start with the conventional form for the differential scattering cross section of RXS \cite{blume,ishihara0}
\begin{equation}
{d^2 \sigma \over d \Omega d \omega_f}= A {\omega_f \over \omega_i}  \sum_{| f \rangle }|S|^2 
 \delta(\varepsilon_f+\omega_f-\varepsilon_i-\omega_i) , 
 \label{eq:sigma}
\end{equation}
where 
\begin{eqnarray}
S&=&  \sum_m \Biggl \{ 
{ \langle f | \vec j_{-k_i} \cdot \vec e_{k_i \lambda_i }|m \rangle 
  \langle m | \vec j_{ k_f} \cdot \vec e_{k_f \lambda_f}  |i \rangle 
  \over \varepsilon_i-\varepsilon_m-\omega_f}
\nonumber \\
& &\ \ \ \ \ 
+
{\langle f | \vec j_{k_f}  \cdot \vec e_{k_f \lambda_f} | m \rangle 
 \langle m | \vec j_{-k_i} \cdot \vec e_{k_i \lambda_i}  | i \rangle 
 \over \varepsilon_i-\varepsilon_m+\omega_i+i \Gamma}
 \Biggr \} ,
 \label{eq:s}  
\end{eqnarray}
with the cross section of the Thomson scattering $A=(e^2/mc^2)^2$. 
$\Gamma$ denotes the damping of a core hole, 
$\vec e_{ k \lambda}$ is the polarization vector of x ray and 
$\vec j_k$ is the current operator.
This form is rewritten by using 
the correlation function of the polarizability $\alpha_{l \beta \alpha }$ 
as shown in Ref.~\onlinecite{ishihara2}: 
\begin{eqnarray}
{d^2 \sigma \over d \Omega d \omega_f}=A {\omega_f \over \omega_i}
\sum_{\alpha \beta \alpha' \beta'}
P_{\beta' \alpha' } P_{ \beta \alpha}
\Pi_{\beta' \alpha' \beta \alpha}(\omega, \vec K) , 
\label{eq:sigma2}
\end{eqnarray}
where
%
%
%figure 5
\begin{figure}
\epsfxsize=0.8\columnwidth
\centerline{\epsffile{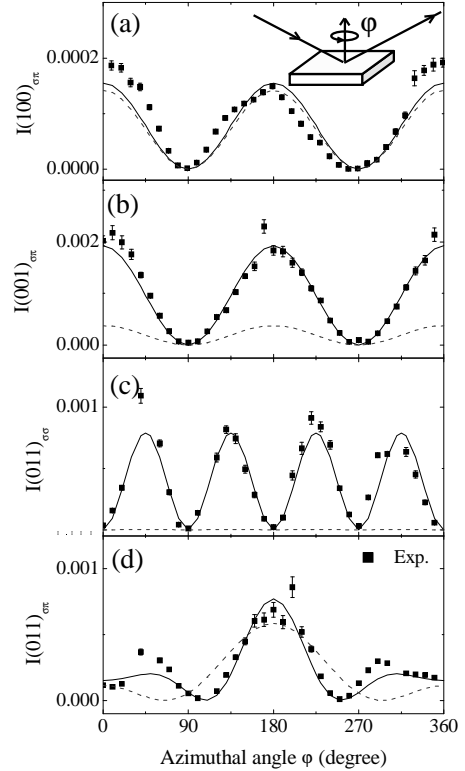}}
\caption{
The azimuthal angle dependence of the resonant x-ray scattering intensity. 
(a) $I(100)_{\sigma \pi}$, (b) $I(001)_{\sigma \pi}$, 
(c) $I(011)_{\sigma \sigma}$ and (d) $I(011)_{\sigma \pi}$. 
The bold curves show the fitted results by 
the wave functions $(a,b,c)=(-0.71, 0, 0.71)$. 
The doted curves show the calculated results by  
the wave functions 
$(a,b,c)=(1/\sqrt{2},0,i /\sqrt{2})$. 
The filled squares indicate the experimental data in YTiO$_3$ 
normalized by the intensity at (022) obtained in Ref.~\protect\onlinecite{nakao}. 
The inset of (a) shows a schematic picture of experimental arrangement. 
}
\label{fig:rxs}
\end{figure}
\begin{eqnarray}
\Pi_{\beta' \alpha' \beta \alpha}(\omega, \vec K)&=&{1 \over 2 \pi}
\int dt e^{i \omega t} \sum_{ll'} e^{-i\vec K \cdot (\vec r_{l'}-\vec r_{l})}
\nonumber \\
&\times&
\langle i |\alpha_{l' \beta' \alpha'}(t)^\dagger 
\alpha_{l \beta \alpha}(0)| i  \rangle , 
\label{eq:pialal}
\end{eqnarray}
with 
$\vec K=\vec k_i-\vec k_f$, $\omega=\omega_i-\omega_f$ 
and $P_{\beta \alpha }=(\vec e_{ k_f \lambda_f})_{\beta}   (\vec e_{k_i \lambda_i})_{\alpha}$. 
Now we express the polarizability operator by the orbital operators $O_{l \Gamma \gamma}$ 
by utilizing the group theoretical analyses.  
$\alpha_{l \beta \alpha }$ associated with the orbital operators at site $l$ 
is represented by 
\begin{equation}
\alpha_{l \beta \alpha }(t)=\sum_{\Gamma \gamma} I_{\Gamma} (M_{\Gamma \gamma})_{ \beta \alpha} 
O_{l \Gamma \gamma}(t) , 
\label{eq:expand}
\end{equation} 
with $(\Gamma \gamma)=(Eu, Ev)$, $(T_1x,T_1y,T_1z)$ and $(T_2x,T_2y,T_2z)$.
$I_{\Gamma}$'s are coupling constants which are not determined by the group theoretical analyses 
and $M_{\Gamma \gamma}$'s are matrices with respect to the polarization of x ray. 
Explicit forms of $M_{\Gamma \gamma}$ are given by 
the Gell-Mann matrices $\lambda_l$ introduced in Sec.~\ref{sec:hamiltonian} 
as
$M_{\Gamma \gamma}=\lambda_l$ for 
$(\Gamma \gamma; l)=(Eu; 8)$, $(Ev, 3)$, 
$M_{\Gamma \gamma}=-\lambda_l$ for 
$(T_2 x; 6)$, $(T_2 y; 4)$, $(T_2 z; 1)$, and 
$M_{\Gamma \gamma}=-i\lambda_l$ for 
$(T_1 x; 7)$, $(T_1 y; 5)$, $(T_1 z; 2)$. 
As a result, the cross section for the static scattering 
in the orbital ordered state is obtained as
\begin{equation}
{d \sigma \over d \Omega} = {A \over 2 \pi}N^2
\bigl |B(\vec K) \bigr |^2 , 
\label{eq:static}
\end{equation}
with 
\begin{equation}
B(\vec K)=\sum_{\Gamma \gamma}  I_{\Gamma \gamma}
S_{\Gamma \gamma}
\langle O_{\Gamma \gamma}(\vec K) \rangle . 
\label{eq:bbbb}
\end{equation}
Here, 
$S_{\Gamma \gamma}$ is the polarization part of the cross section defined by 
\begin{equation}
S_{\Gamma \gamma}=\vec e_{ k_f \lambda_f}^{\ t} M_{\Gamma \gamma} \vec e_{ k_i \lambda_i} , 
\label{eq:structure}
\end{equation}
and 
$ \langle O_{\Gamma \gamma}(\vec K) \rangle$ is the orbital order parameter 
\begin{equation}
 \langle O_{\Gamma \gamma}(\vec K) \rangle=
{1 \over N}\sum_{l} e^{i \vec K \cdot \vec r_l} 
\langle O_{l \Gamma \gamma} \rangle , 
\end{equation}
with a number of the unit cell $N$. 
Equation~(\ref{eq:static}) with Eq.~(\ref{eq:bbbb}) is derived by only considering 
the symmetries of crystal and orbital. 
Therefore, we do not specify, in this paper, 
the microscopic origin of the orbital dependence of the polarizability. 
However, it is reported recently that 
the mechanism of RXS based on the Coulomb interactions between $3d$ and $4p$ electrons 
well explains relative scattering intensities at different reflection points 
in YTiO$_3$, rather than the mechanism based on the lattice distortions. \cite{nakao,takahashi}

The above results are applied to 
RXS in YTiO$_3$ where the detailed experiments have been done recently 
and reported in Ref.~\onlinecite{nakao}. 
In the RXS experiments, the azimuthal angle scan, which is the rotational scan about 
the scattering vector $\vec K$, is crucially important to identify the orbital ordering. \cite{murakami} 
This is because the local symmetry of the orbital structure directly 
reflects on the azimuthal angle dependence. 
A schematic picture of the experimental arrangement is presented in the inset of Fig.~\ref{fig:rxs} (a).  
In this arrangement, $S_{\Gamma \gamma}$ in Eq.~(\ref{eq:structure}) is replaced by
\begin{equation}
S_{\Gamma \gamma}=\vec e_{ k_f \lambda_f}^{\ t} U (\varphi) V 
M_{\Gamma \gamma} V^{-1} U^{-1}(\varphi) \vec e_{ k_i \lambda_i} , 
\label{eq:structure2}
\end{equation}
where the matrix $V$ describes the transformation   
from the crystallographic coordinate to the laboratory coordinate 
and the matrix $U(\varphi)$ represents the azimuthal rotation with the rotation angle $\varphi$. 
The GdFeO$_3$-type lattice distortion is also taken into account.
A general formalism for the azimuthal angle dependence of RXS intensity 
is presented in Ref.~\onlinecite{ishihara3}. 
The RXS experiments have been carried out at three orbital superlattice reflection points 
of $(100)$, $(001)$ and $(011)$, where we use the $Pbnm$ orthorhombic notations. 
We fit the four sets of the experimental data: 
$(100)$ with $(\lambda_i, \lambda_f)=(\sigma,\pi)$, 
$(001)$ with $(\sigma, \pi$), 
$(011)$ with $(\sigma, \sigma)$, 
and 
$(011)$ with $(\sigma, \pi)$, 
where $\sigma$ ($\pi$) indicates the $\sigma$ ($\pi$) polarization of x ray. 
From now on, 
the scattering intensity at $(hkl)$ with polarizations $(\lambda_i,\lambda_f)$ 
is denoted by $I(hkl)_{\lambda_i \lambda_f}$. 
$I(100)_{\sigma \sigma}$ and $I(001)_{\sigma \sigma}$ are zero within the experimental errors.
We assume that 
the orbital wave functions have a symmetry of the $Pbnm$ point group. 
Then, the coefficients $C_{i \alpha}$ in the orbital wave functions (see Eq.~(\ref{eq:ws})) 
satisfy the conditions 
$C_{A_1 yz}= C_{A_2 yz}= C_{B_1 zx}=C_{B_2 zx} \equiv a$, 
$C_{A_1 zx}= C_{A_2 zx}= C_{B_1 yz}=C_{B_2 yz} \equiv b$, 
and 
$C_{A_1 xy}=-C_{A_2 xy}=C_{B_1 xy}=-C_{B_2 xy} \equiv c$.  
In the case where $a$, $b$ and $c$ are real, 
the explicit forms of the scattering intensities are given by 
\begin{eqnarray}
I(100)_{\sigma \sigma}&=& 0 , \\
I(100)_{\sigma \pi}&=& I_0  (4I_{E}  O_{Ev} \sin \varphi \cos \theta)^2 ,\\
I(001)_{\sigma \sigma}&=& 0 , \\
I(001)_{\sigma \pi}&=& I_0 (4I_{T_2} O_{T_2} \sin \varphi \cos \theta)^2 , \\
I(011)_{\sigma \sigma}&=& I_0 8 (I_{T_2} O_{T_2} \sin 2\varphi)^2 , \\
I(011)_{\sigma \pi}&=& I_0 8 (I_{T_2} O_{T_2})^2 \nonumber \\ 
  &\times& (-\cos 2\varphi \sin \theta+\sin \varphi \cos \theta)^2 , 
\end{eqnarray}
where the GaFeO$_3$-type distortion is neglected.
$\theta$ is the scattering angle, $I_0=AN^2/(2 \pi)$ and, 
$O_{Ev}$ and $O_{T_2}$ are the order parameters given by 
\begin{equation}
O_{Ev}   = \langle O_{A_1 Ev} \rangle={1 \over \sqrt{2} }(a^2-b^2) ,
\end{equation}and
\begin{equation}
O_{T_2}={1 \over 2} (\langle O_{A_1 T_2 x} \rangle-\langle O_{A_2 T_2 y}\rangle)  = {1 \over \sqrt{2}} c(a-b) , 
\end{equation}
respectively. 
It is worth mentioning that $I(100)_{\sigma \pi}$ and other $I$'s are reflected 
from different components of the orbital order parameters, 
i.e. $O_{Ev}$ and $O_{T_2}$, respectively. 
From the above considerations, 
$a$, $b$ and $c$ are restricted so that 
$c$ is finite and $a \ne b$. 
We optimize values of $a$, $b$, $c$ and $I_{T_2}/I_{E}$ numerically 
within real numbers.
It is found that the calculated results fit in the four sets of the experimental data simultaneously, 
when the wave function $(a,b,c)$ satisfies the condition 
$c \geqap -a >> |b| $. 
One of the best fitted results is obtained by 
a set of the parameters 
$(a,b,c)=(-0.71, 0, 0.71)$, 
being consistent with the recent experimental analyses,\cite{nakao,akimitsu} 
and $I_{T_2}/I_E=0.45$. 
The results are shown in Fig.~\ref{fig:rxs} (bold lines) 
together with the experimental data (filled squares). 
The fitting by the calculation is satisfactory. 
For comparison, we calculate the RXS intensity 
where the orbital wave functions are complex (dotted lines in Fig.~\ref{fig:rxs}). 
We set $a=1/\sqrt{2}$, $b=0$ and $c=i/\sqrt{2}$. In this case, 
the scattering intensities are explicitly given by 
\begin{eqnarray}
I(100)_{\sigma \pi}   &=& I_0 (4I_{E} O_{Ev}    \sin \varphi \cos \theta)^2,   \\
I(001)_{\sigma \pi}   &=& I_0 ( I_{T_1} O_{T_1} \sin \varphi \cos \theta)^2,  \\  
I(011)_{\sigma \pi}   &=& I_0 {1 \over 2}  (I_{T_1} O_{T_1 })^2 \nonumber \\
&\times& (\sin \theta+\sin \varphi \cos \theta)^2 , 
\end{eqnarray}
and $I_{\sigma \sigma}$'s are zero.  
$O_{T_1}$ is the order parameter for the magnetic dipole moment defined by 
\begin{eqnarray}
O_{T_1}    &=& \langle O_{A_1 T_1 x}   \rangle={i \over \sqrt{2}} (a^\ast c-c^\ast a) . 
\end{eqnarray}
The discrepancies between the calculated results and the experimental data are remarkable, 
especially, in $I(011)_{\sigma \sigma}$.  
This is because the order parameter $O_{T_2}$ vanishes  
and $O_{T_1}$ appears in this orbital ordered state. 

\begin{figure}
\epsfxsize=0.8 \columnwidth
\centerline{\epsffile{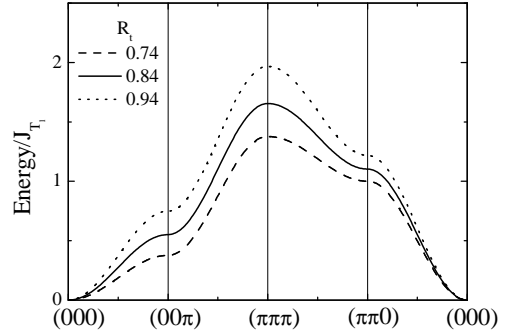}}
\caption{
The dispersion relation of the spin wave.
Broken, bold and dotted curves are the results with 
$R_t$=0.74, 0.84 and 0.94, respectively. 
The cubic notation is used. 
The orbital wave functions are chosen as 
$(a,b,c)=(-0.71, 0, 0.71)$, and $R_J$ is fixed to be 0.25.  
The ratio of the spin wave stiffness constant $R_D=D_{zz}/D_{xy}$ 
and the Curie temperature $T_C$ is  
$(R_D,T_C/J_{T_1})=(0.75, 0.21), (1.00, 0.25), (1.23, 0.30)$ 
for $R_t$=0.74, 0.84 and 0.94, respectively. 
}
\label{fig:sw}
\end{figure}

By using the orbital wave function 
$(a,b,c)=(-0.71, 0, 0.71)$ obtained above, 
the spin wave dispersion relation is calculated by applying the 
Holstein-Primakoff transformation.  
The results are presented in Fig.~\ref{fig:sw} for several values of $R_t$. 
In the same sets of parameters, 
the Curie temperature $T_C$ is calculated by the mean-field approximation 
and calculated values are corrected by considering 
the results in the high temperature expansion. \cite{domb}
$R_D=D_{zz}/D_{xy}$, where $D_{zz}$ ($D_{xy}$) is the spin stiffness in the $z$ direction (the $xy$ plane), 
and $T_C$ is obtained as 
$(R_D,T_C/J_{T_1})=(0.75, 0.21), (1.00, 0.25), (1.23, 0.30)$ 
for $R_t=$0.74, 0.84, 0.94, respectively. 
The spin wave in YTiO$_3$ 
is recently measured in Ref.~\onlinecite{keimer2} by neutron scattering experiments 
and its dispersion relation is found to be almost isotropic. 
A value of $J_{T_1}$ is estimated by fitting the experimental data 
by the calculated results of $R_t=0.84$ 
where the spin stiffness is almost isotropic. 
The obtained value is $J_{T_1}=10.9$meV by which 
$T_C$ is given by 32K. 
This result is consistent with $T_C$ in YTiO$_3$ of about 30K. 
%
%figure6

\section{summary}

In summary, 
we derive the effective Hamiltonian for spin and orbital states in perovskite 
titanates $R_{1-x}A_x$TiO$_3$. 
By taking into account the three-fold orbital degeneracy, 
the orbital parts of the Hamiltonian are represented by 
the eight $3 \times 3$ matrices.  
The orbital ordered states in the ferromagnetic phase for the end compounds have 
strong continuous degeneracy. 
This is owing to the three-fold degenerated $t_{2g}$ orbitals 
and the orthogonality of the electron hopping integral between NN ions, 
unlike the two-fold degenerated $e_g$ orbitals. 
Introductions of the GdFeO$_3$-type distortion, the JT-type distortion 
and the LS coupling lift the orbital degeneracy and destabilize 
the ferromagnetic state. 
This implies that the high symmetry in the orbital space is favored for the ferromagnetic ordering. 
The scattering cross section of RXS is formulated by utilizing 
the same orbital operators adopted in the Hamiltonian. 
It is shown that the different components of the orbital order parameters 
are detected separately at the different reflection points and polarization configurations.  
The experimental data of the azimuthal angle dependent RXS intensities are well fitted by 
the present calculation.  

The present theory based on the derived effective Hamiltonian and new RXS formula 
is satisfactory to explain 
the several experiments about spin and orbital states in YTiO$_3$; 
the ferromagnetic structure, the azimuthal angle dependence of the RXS intensity, 
the Curie temperature and the spin wave dispersion relation. 
It is thought that the spin and orbital states are controlled by the interactions 
between NN Ti sites through the virtual electron exchange processes  
and the associated JT-type lattice distortions. 
The present RXS studies suggest that the LS coupling may be irrelevant and 
the real orbital wave functions are realized by the inter-site interaction between orbitals. 

Beyond the static spin and orbital ordered states in $R$TiO$_3$, 
this Hamiltonian is applicable to the orbital excitations termed orbital wave \cite{ishihara}
and to the hole doped systems $R_{1-x}A_{x}$TiO$_3$. 
For the actual calculations, 
the present Hamiltonian represented by the $3 \times 3$ matrices $O_{i \Gamma \gamma}$ 
is more convenient than that by the $ 2 \times 2$ matrices $T_{i l}$ (see Eq.~(\ref{eq:pspsps})) 
with the constraint and the Hubbard type Hamiltonian. 
The study of the doped titanates and the orbital dynamics will be presented in a separate publication. 

\begin{acknowledgments}
Authors would like to thank Y.~Murakami, H.~Nakao, D.~Gibbs and J.~P.~Hill 
for providing their experimental data prior publication. 
Authors would appreciate G.~Khaliullin and B.~Keimer for their valuable discussions. 
This work was supported by the Grant in Aid from Ministry of Education, Culture, 
Sports, Science and Technology of Japan, CREST, and Science and Technology Special 
Coordination Fund for Promoting Science and Technology. 
One of authors (S.M.) acknowledges support of the Hunboldt Foundation. 
\end{acknowledgments}
\appendix
\section{Effective Hamiltonian with a GdFeO$_3$-type lattice distortion}
Effects of the GdFeO$_3$-type lattice distortion are included in 
the modified $J$ term ${\widetilde {\cal H}}_J$
introduced in Eq.~(\ref{eq:effective2}).  
The explicit form of this term is given by 
\begin{equation}
{\widetilde {\cal H}}_J={\cal H}_J
                       +{\widetilde {\cal H}}_{T_1}+{\widetilde {\cal H}}_{T_2}
                       +{\widetilde {\cal H}}_{E}+{\widetilde {\cal H}}_{A_1}, 
\label{eq:eff2}
\end{equation}
with 
\label{sec:ap1}
\begin{eqnarray}
{\widetilde{\cal H}}_{T_1}&=&-J_{T_1} \sum_{\langle ij \rangle}
\biggl ( {3 \over 4}n_i n_j+\vec S_i \cdot \vec S_j \biggr) 
\nonumber \\
&\times& 
\biggl \{ s_l R_t 2D^{'l}+R_t^2 \biggl ({1 \over 2}A^l-C^{'l}+D^l \biggr )   \biggr \} , 
\label{eq:ht1b}
\end{eqnarray}
\begin{eqnarray}
{\widetilde{\cal H}}_{T_2}&=&-J_{T_2} \sum_{\langle ij \rangle}
\biggl ( {1 \over 4} n_i n_j-\vec S_i \cdot \vec S_j \biggr) 
\nonumber \\
&\times& 
\biggl \{ s_l R_t 2(D^{'l}+E^{'l})+R_t ^2 \biggl ({1 \over 2}A^l+C^{'l}+D^l \biggr )  \biggr \} , 
\label{eq:ht2b}
\end{eqnarray} 
\begin{eqnarray}
{\widetilde{\cal H}}_{E}&=&-J_{E} \sum_{\langle ij \rangle}
\biggl ( {1 \over 4}n_i n_j-\vec S_i \cdot \vec S_j \biggr) \nonumber \\
&\times &
\biggl \{ s_l R_t {2 \over 3} E^l+R_t^2 \biggl ({4 \over 3} B^l -{2 \over 3} C^l\biggr ) \biggr \}  , 
\label{eq:heb}
\end{eqnarray}
\begin{eqnarray}
{\widetilde {\cal H}}_{A_1}&=&-J_{A_1} \sum_{\langle ij \rangle}
\biggl ( {1 \over 4}n_i n_j-\vec S_i \cdot \vec S_j \biggr) \nonumber \\
& \times & 
\biggl \{ s_l R_t{4 \over 3}E^l+R_t^2 \biggl ({2 \over 3} B^l +{2 \over 3}c^l \biggr ) \biggr \}   . 
\label{eq:ha1b}
\end{eqnarray}
The parameter $R_t=t_\sigma/t_\pi$ is caused by the 
GdFeO$_3$-type distortion. 
$A^l$, $B^l$, $C^l$, $C^{'l}$ and $D^l$ are defined in Eqs.~(\ref{eq:aaa}), (\ref{eq:bbb}), 
(\ref{eq:ccc}), (\ref{eq:ccc2}) and (\ref{eq:ddd}), respectively. 
$D^{'l}$ and $E^l$ 
describe new exchange processes induced by the distortion 
and are given by 
\begin{eqnarray}
D^{'l}=&-& \sqrt{2} 
     \biggl( {1 \over 3} + \sqrt{2 \over 3} O^l_{i Eu} \biggr )
     O_{j T_2 l }
     \nonumber \\
     &-& \sqrt{2} 
     O_{i T_2 l }
     \biggl( {1 \over 3} + \sqrt{2 \over 3} O^l_{j Eu} \biggr ) , 
\label{eq:ddd2}
\end{eqnarray}
\begin{eqnarray}
E^l=&-& \sqrt{2} 
     \biggl( {2 \over 3} -\sqrt{2\over 3} O^l_{i Eu} \biggr )
     O_{j T_2 l }
     \nonumber \\
    &-&\sqrt{2}
     O_{i T_2 l }
     \biggl( {2 \over 3} -\sqrt{2 \over 3} O^l_{j Eu} \biggr ) . 
\label{eq:eee}
\end{eqnarray}
By utilizing the pseudo-spin operator introduced in Eq.~(\ref{eq:pspsps}), 
these terms are rewritten as 

\begin{equation}
D^{'l}=2(n_{i c_l}T_{jx}^l+T_{ix}^l n_{j c_l}) , 
\end{equation}
\begin{eqnarray}
E^{l}=2 \bigl \{ (n_{i a_l}+n_{i b_l})T_{jx}^l+T_{ix}^l(n_{j a_l}+n_{j b_l}) \bigr \} . 
\end{eqnarray}

\end{document}